# Lattice tuning of optical and mechanical interactions in gold nanodisk metasurfaces

*J. I. Sangiorgio*[1,2,†], *D. Xie*[3,†], *E. Cortés*[3], *G. Grinblat*[1,2] *and A. V. Bragas*[1,2*]

1-Universidad de Buenos Aires, Facultad de Ciencias Exactas y Naturales, Departamento de Física, Buenos Aires 1428, Argentina;

2-CONICET-Universidad de Buenos Aires, Instituto de Física de Buenos Aires (IFIBA), Buenos Aires 1428, Argentina;

3- Nanoinstitute Munich, Faculty of Physics, Ludwig-Maximilians-Universität München, Munich, Germany

†-These authors contributed equally.

*Corresponding author: bragas@df.uba.ar

# Abstract

The interaction between neighboring plasmonic nanoparticles can modify their intrinsic optical and mechanical resonances, enabling collective responses to be controlled through the geometry of their arrangement. Periodic arrays therefore provide a means of controlling both electromagnetic and mechanical interactions through a common geometrical parameter, the lattice pitch. Joint control of these two channels is particularly useful for engineering how coherent mechanical excitations couple to the substrate and are optically read out at the nanoscale. Here, we investigate the optical and mechanical response of gold nanodisk metasurface patches on silica as a function of lattice pitch using two-color pump–probe spectroscopy and numerical simulations. Optically, simulations show that varying the pitch tunes a Rayleigh anomaly across the broad localized surface plasmon resonance of the nanodisks, producing a strongly asymmetric collective optical response. Pump–probe measurements further reveal a pronounced wavelength dependence of the optical readout of coherent mechanical vibrations. Mechanically, we investigate how coupling between the nanodisks and the substrate depends on the character of the nanodisk resonances. The intrinsic nanodisk modes exhibit markedly different sensitivities to the periodic environment. One remains nearly independent of pitch, whereas another undergoes a pronounced frequency shift mediated by elastic interactions through the substrate. Numerical simulations further reveal that a third nanodisk resonance couples resonantly to a lattice-selected acoustic mode, giving rise to an avoided crossing and mode hybridization. These distinct mechanical responses are governed by how efficiently each mode couples to near-surface elastic motion in the substrate. These results demonstrate that lattice periodicity can simultaneously tune optical sensitivity and substrate-mediated mechanical coupling, providing a versatile platform for tailoring nanoscale optomechanical responses in the GHz regime.

# Introduction

Plasmonic nanostructures have the ability to confine and enhance electromagnetic fields in sub-wavelength volumes.[1–4] Owing to this unique behavior, they have had a strong impact on a wide range of areas, such as biosensing,[5–7] cancer therapy,[8,9] surface-enhanced Raman spectroscopy,[10,11] and catalysis,[12] to name a few. In addition to their optical response, plasmonic nanoparticles support coherent mechanical vibrations in the gigahertz range,[13,14] making them promising candidates for optomechanical signal transduction at the nanoscale.[15,16] For an isolated nanoantenna on a solid substrate, acoustic radiation into the substrate constitutes a loss channel, contributing to mechanical damping and limiting the quality factor.[15,17] Once arranged periodically, however, the elastic response of the individual elements becomes coupled through the substrate and constrained by the lattice periodicity, giving rise to collective mechanical modes.[18,19]

Metasurfaces, both dielectric and plasmonic, have been extensively explored for a broad range of applications, including flat optics,[20,21] beam steering,[22,23] nonlinear optics,[24–26] optical communications,[27,28] sensing,[29,30] energy,[31] and imaging.[32,33] Periodicity also plays a central role in the optical response of plasmonic metasurfaces. Localized surface plasmon resonances (LSPR) can interact with diffractive lattice modes, giving rise to surface lattice resonances.[34,35] Thus, a periodic plasmonic array provides a natural platform in which the same geometric parameter, the lattice pitch, can organize collective responses in two different physical domains, through electromagnetic coupling mediated by diffraction and elastic coupling mediated by acoustic waves in the substrate. Controlling both interactions within the same structure is therefore attractive because the optical response governs the sensitivity to coherent vibrations, while substrate-mediated elastic coupling determines how mechanical energy is distributed between the nanodisks and the substrate. While collective optical effects in periodic plasmonic arrays are well established, the way periodicity reshapes their mechanical mode structure, particularly through the interaction between localized nanodisk resonances and substrate-mediated lattice modes, remains much less explored.

In this work, we use two-color pump-probe spectroscopy to investigate the mechanical response of gold nanodisk metasurfaces on silica. A central aspect of our study is the influence of lattice pitch on this mechanical response. By combining pump-probe measurements with numerical simulations, we track the dependence of the intrinsic nanodisk resonances and collective acoustic lattice modes on lattice pitch, and investigate how their frequencies and modal character are modified by substrate-mediated interactions. This analysis provides insight into the coupling between localized mechanical resonances and lattice modes, revealing distinct responses including weakly pitch-dependent intrinsic modes, substrate-mediated frequency shifts, dispersive lattice-selected acoustic modes, and their resonant hybridization. These different responses provide a means of controlling not only the frequencies of the mechanical excitations, but also their spatial distribution and coupling to the substrate. In parallel, tuning the lattice pitch shifts the relevant diffractive condition across the broad localized surface plasmon resonance of the gold nanodisks, producing a pronounced asymmetric optical response.[34–36] The lattice pitch therefore provides a common geometrical parameter to tune both optical and mechanical collective responses. Overall, we propose a tunable, easy-to-fabricate optomechanical platform for controlling and probing both the collective mechanical excitations in the GHz range, and the spatial distribution of optical and mechanical modes, with potential for photonic-phononic signal processing.

# Results and discussion

Gold nanodisks were fabricated on a $SiO_2$ substrate by electron-beam lithography in two configurations: isolated disks and periodic metasurfaces (Fig. 1). The isolated-disk structures were patterned with a 3 μm pitch (p), sufficiently large to suppress coupling between neighbors. The metasurface patches consisted of finite six-by-six square arrays, with pitches ranging from 400 nm to 500 nm. All disks had a nominal diameter of 150 nm and a height of 35 nm, chosen to place the localized plasmonic response close to the ~800 nm probe wavelength.

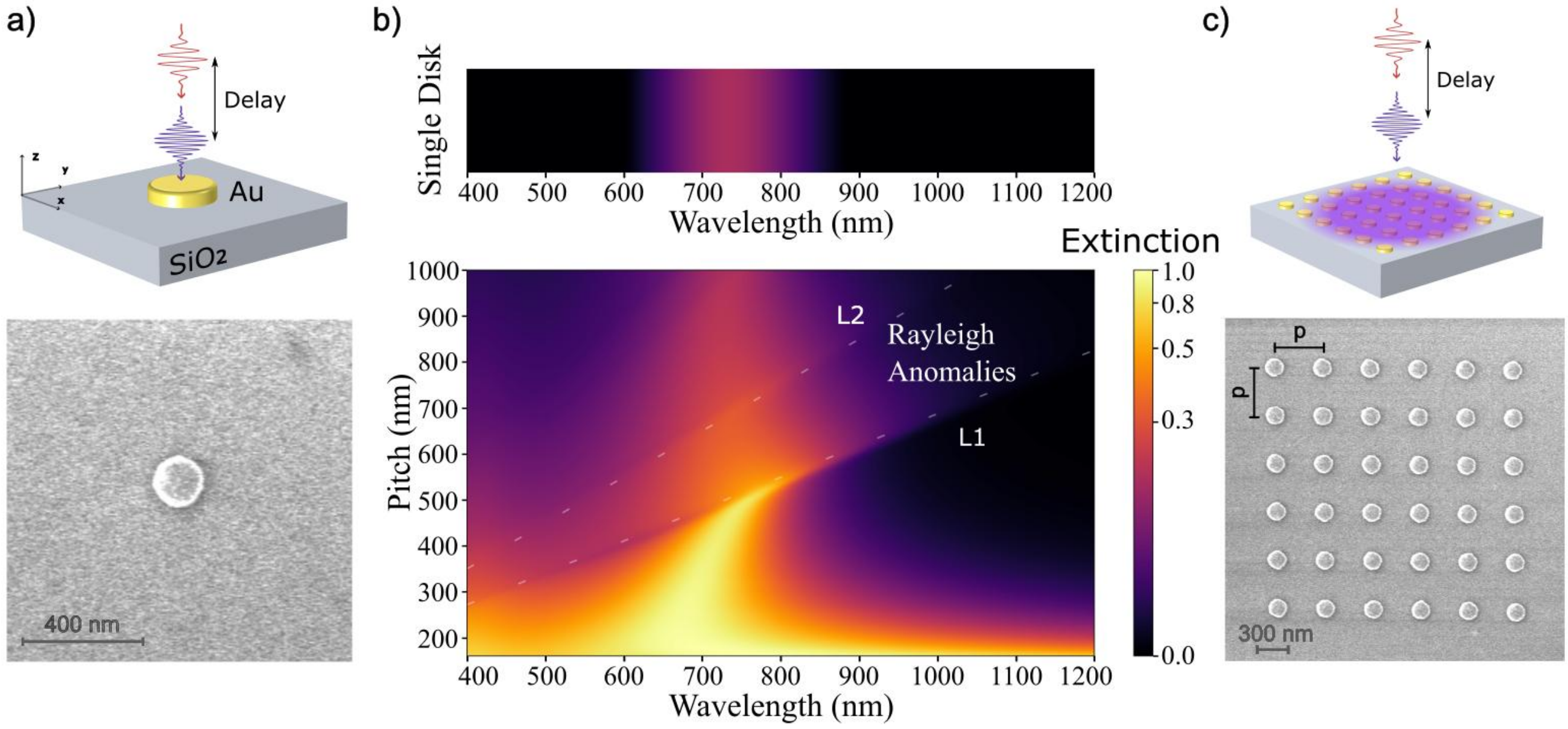

Figure 1. (a) Representation of the individual nanodisk system along with its corresponding scanning electron microscopy (SEM) image. The mechanical response of gold nanodisks was analyzed using two-color pump–probe spectroscopy. (b) Top: Extinction cross-section for a single nanodisk, showing that the optical response is dominated by the plasmonic resonance. Bottom: Extinction cross-section for a gold nanodisk metasurface as a function of wavelength and pitch. Dashed lines L1 and L2 correspond to expected Rayleigh anomalies. (c) Schematic of a gold metasurface patch indicating the six-by-six array being pumped collectively with its corresponding SEM image.

The optical and mechanical response of these systems was studied using two-color pump-probe spectroscopy. A 400 nm pump impulsively heats the electronic subsystem of the individual gold nanodisks, and the subsequent electron-lattice thermalization causes a rapid thermoelastic expansion that launches coherent mechanical vibrations. These deformations modulate the optical response, producing a measurable change in transmission that is detected by the 800 nm probe pulse at a controlled time delay.

## Optical response and optomechanical readout

Since the sensitivity of our experiment strongly depends on the optical response of the sample, we first characterize the extinction of the individual and periodic structures. Figure 1.b (top) shows the extinction cross-section for an isolated disk, dominated by the broad plasmonic response around 800 nm. In the metasurfaces, where the disk separation becomes comparable to the incident wavelength, electromagnetic interactions between nanodisks modify this response. Figure 1.b (bottom) shows the calculated extinction as a function of the wavelength of incident light, and pitch, where brighter (yellow) areas correspond to higher extinction. The colormap shows both the evolution of the broad plasmonic resonance and additional dispersive features arising from the periodic arrangement.

In addition to the broad localized plasmonic response, the extinction map displays diagonal features that disperse with the lattice pitch. The gray dashed lines indicate the Rayleigh anomalies (RA) predicted by the empty-lattice approximation for the two lower diffraction orders, L1 and L2 modes. A Rayleigh anomaly marks the wavelength at which a diffraction order becomes grazing and is therefore determined by the lattice geometry and refractive environment, independently of the plasmonic properties of the individual nanodisks.[34,35] The collective lattice resonance (LR) results from the interplay between the individual nanodisk polarizability and the electromagnetic interaction mediated by the periodic array. Within a dipolar description, this response can be expressed through an effective array polarizability $\alpha_{eff} = [\alpha^{-1} - G]^{-1}$, where $\alpha$ describes the single-particle response and $G$ is the lattice sum.[37,38] The lattice resonance occurs approximately when $Re[\alpha^{-1} - G] = 0$[38] and is therefore generally spectrally displaced from the corresponding Rayleigh anomaly. In our metasurface the first Rayleigh anomaly is tuned across the broad LSPR of the individual gold nanodisks as the pitch is varied. Thus, unlike the case in which the single-particle polarizability varies only weakly across the lattice resonance, here both the localized plasmonic response and the lattice-mediated interaction vary significantly over the same spectral range. Their spectral overlap strongly reshapes the extinction spectrum and produces the asymmetric, Fano-like lineshape observed in Fig. 1b.[34–36]

Figure 2.a illustrates this behavior for p=500 nm by comparing the extinction of the array with that of a single nanodisk. For this pitch, the L1 empty-lattice condition gives $\lambda_{RA}$≈725 nm, whereas the extinction maximum associated with the collective lattice resonance occurs around $\lambda_{LR}$≈760 nm. The comparison with the single-disk spectrum also shows that the Rayleigh anomaly occurs in the spectral region of the broad LSPR, resulting in the pronounced asymmetric lineshape of the array response. Importantly, the probe spectral range lies on the rapidly varying long-wavelength side of this collective optical response. Small changes in extinction induced by the nanodisk vibration can therefore produce large changes in transmission, enhancing the detected amplitude of the coherent mechanical mode. Anticipating the mechanical analysis below, Fig. 2b illustrates this effect, as the probe wavelength is varied the measured phonon amplitude changes markedly with the local slope of the extinction spectrum and more than doubles for a 15 nm change in probe central wavelength.

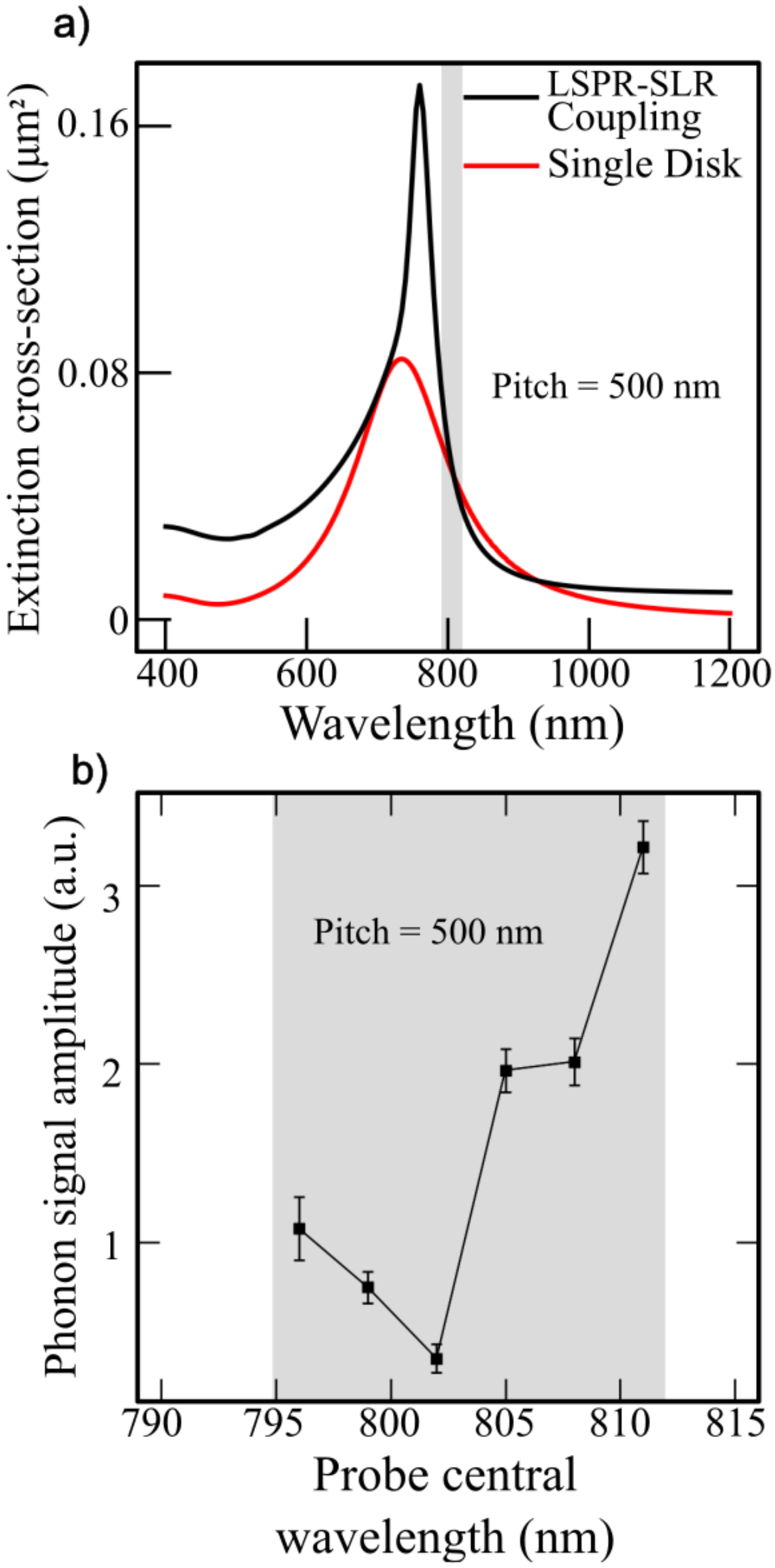

Figure 2. (a) Comparison between the simulated extinction of a single nanodisk (red) and that of a 500 nm-pitch metasurface unit cell. The interplay between the localized plasmon response and the lattice-mediated optical interaction produces the characteristic asymmetric lineshape. The gray area highlights the wavelength range explored experimentally. (b) Measured total amplitude of the phonon signal for a 500 nm-pitch metasurface, showing the strong dependence of the optical readout on probe wavelength.

## Mechanical response of isolated nanodisks

Having established how periodicity reorganizes the electromagnetic response, we now turn to its mechanical counterpart. As a first step, we analyze the vibrational modes of an isolated nanodisk, which provides the reference to understand the collective mechanical behavior that later emerges in the metasurfaces. Figure 3 presents the experimental differential transmission signal obtained on an isolated nanodisk along with its frequency spectrum. A representative pump-probe trace is shown in

Fig. 3a. When the pump and probe pulses arrive simultaneously at the sample, the signal shows a sharp change in transmittance due to the ultrafast electronic response of gold, followed by an exponential decay associated with the thermal relaxation. After subtracting this non-oscillatory background, the phonon signal is extracted and fitted (inset) to obtain its frequency spectrum (Fig. 3b). The experimental spectrum is in good agreement with finite-element simulations of the mechanical response shown in Fig. 4b, which reproduce the observed resonances with only minor frequency deviations. In addition to the spectral response, Fig. 4a shows the simulated displacement patterns of the dominant vibrational modes, including their deformation profiles on the disk and the perturbation induced in the substrate.

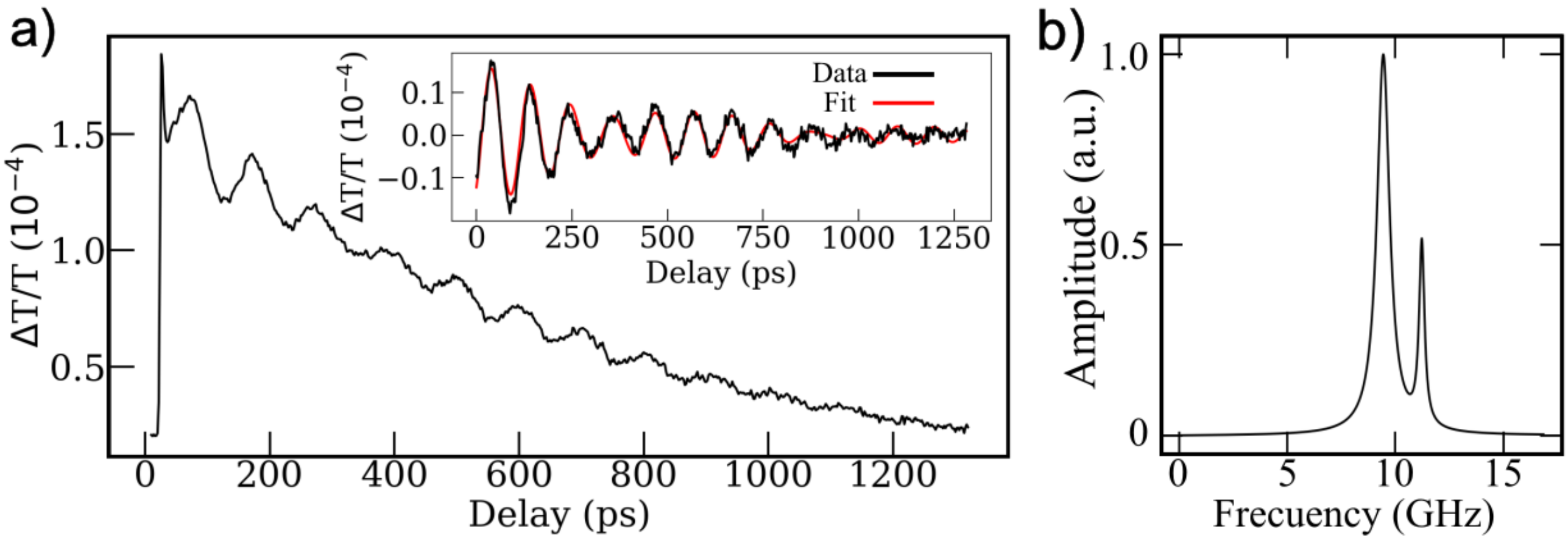

Figure 3. (a) Differential transmission signal of a single nanodisk, showing a strong electronic response followed by thermal relaxation. The inset shows the phonon signal obtained after subtraction of the thermal background and the corresponding fit. (b) Mechanical spectrum obtained from the fit. The extracted modes show good agreement with the dominant resonances predicted by the mechanical simulation in Fig. 4b, apart from a slight frequency mismatch.

For clarity, the three intrinsic nanodisk resonances predicted by the simulations are hereafter denoted modes $n = \text{I}, \text{II}, \text{and III}$ in order of increasing frequency. The displacement fields already reveal a strongly mode-dependent coupling to substrate motion, which will be central to understanding the metasurface response discussed below. To further understand each mode's behavior, Fig. 4b presents the average radial and vertical displacements for an isolated nanodisk, highlighting the difference between the modes. Mode I, for example, is composed almost exclusively of vertical displacement, whereas modes II and III present comparable radial and vertical contributions. Finally, because mechanical coupling between nanodisks is a key property of the metasurfaces, we quantify the fraction of mechanical energy associated with the near-surface region of the $SiO_2$ substrate as $\eta_{SiO_2}(f,p) = \frac{\int_{SiO_2} W\,(\boldsymbol{r};f,p)dV}{\int_{Au+SiO_2} W\,(\boldsymbol{r};f,p)\,dV} = 1 - \eta_{Au}(f,p)$, where $W(\boldsymbol{r};f,p)$ is the energy density at position ***r*** for a metasurface of period p at frequency $f$, with the substrate integral restricted to a depth of one acoustic wavelength. Figure 4c shows that the most prominent intrinsic resonance, mode II, has a comparatively small near-surface substrate contribution, whereas modes I and III couple much more efficiently to surface acoustic motion. The inset shows a representative deformation profile for mode III together with the associated substrate wave.

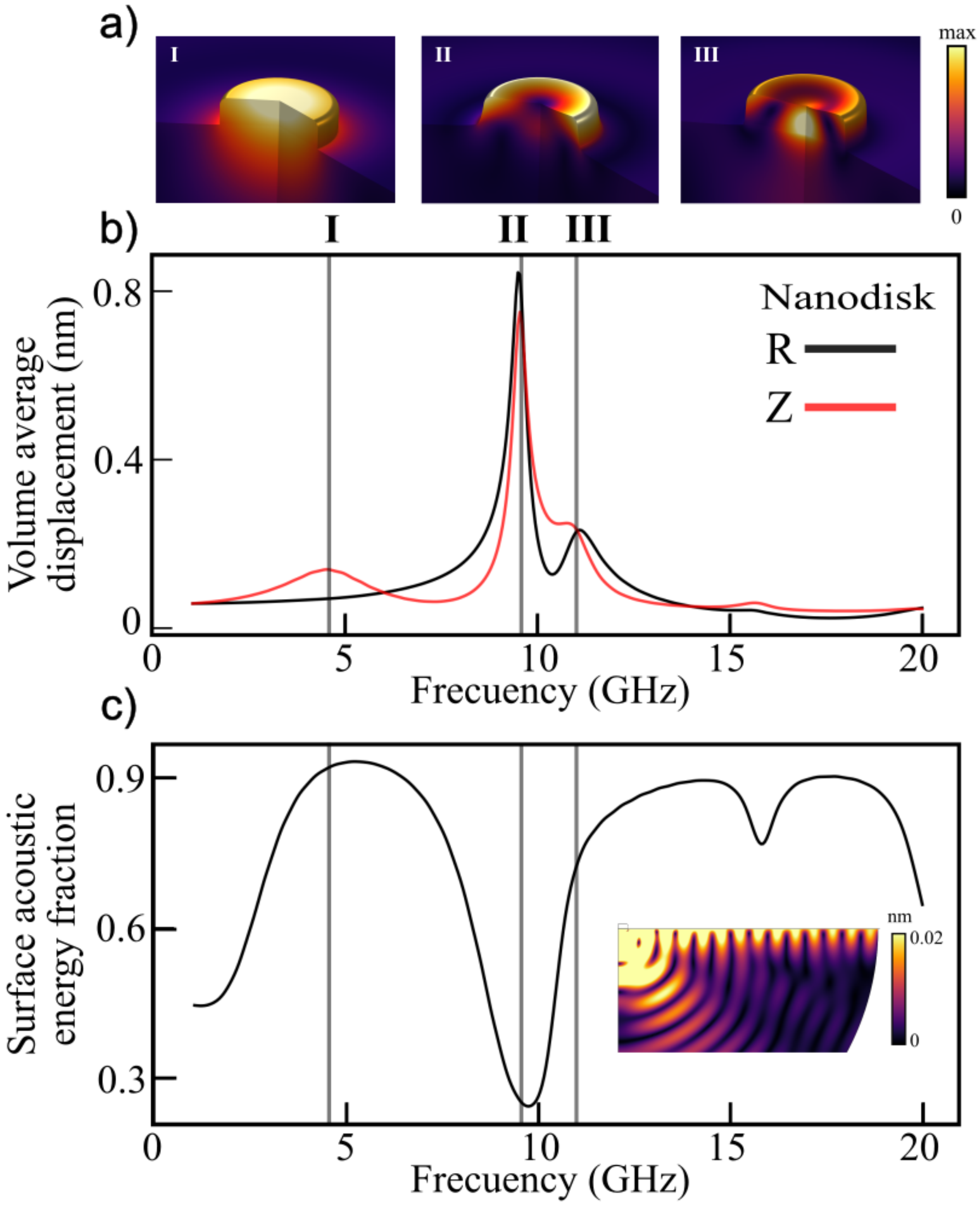

Figure 4. (a) Simulated deformation profiles of the isolated-nanodisk modes I, II, and III marked in (b), showing both the nanodisk displacement and the perturbation induced in the substrate. (b) Mechanical spectrum of the isolated nanodisk, showing the average radial (black) and vertical (red) displacement components of each mode. (c) Surface acoustic energy fraction, quantifying the relative mechanical-energy contribution of the near-surface substrate region (see text). Modes I and III show a substantially higher surface acoustic energy fraction than mode II. The inset shows a representative deformation profile for mode III and the associated surface acoustic wave in the substrate.

## Substrate-mediated mechanical coupling in periodic metasurfaces

As we anticipated, the mechanical response changes when the same nanodisks are arranged in a periodic array. Figure 5.a presents the phonon signal for a 450 nm-pitch metasurface along with its corresponding fit, showing a strong modification in the signal-to-noise ratio compared to the single-disk configuration. This may be partly due to the probe spot effectively averaging the contributions from multiple nanodisks. However, as seen in Figure 5.b, three frequencies are required to account for the mechanical response of the metasurface, compared to two for the single-disk case: two main resonances in the 9–11 GHz range and an additional mode near 5 GHz. This low frequency resonance was already present in the single-disk mechanical simulation, labeled as mode I, although its experimental amplitude was too small to be clearly detected. Its enhanced contribution in the metasurface cannot be explained solely by the larger number of nanodisks probed, since such averaging would improve the overall signal-to-noise ratio without selectively modifying the relative contribution of the different mechanical modes.

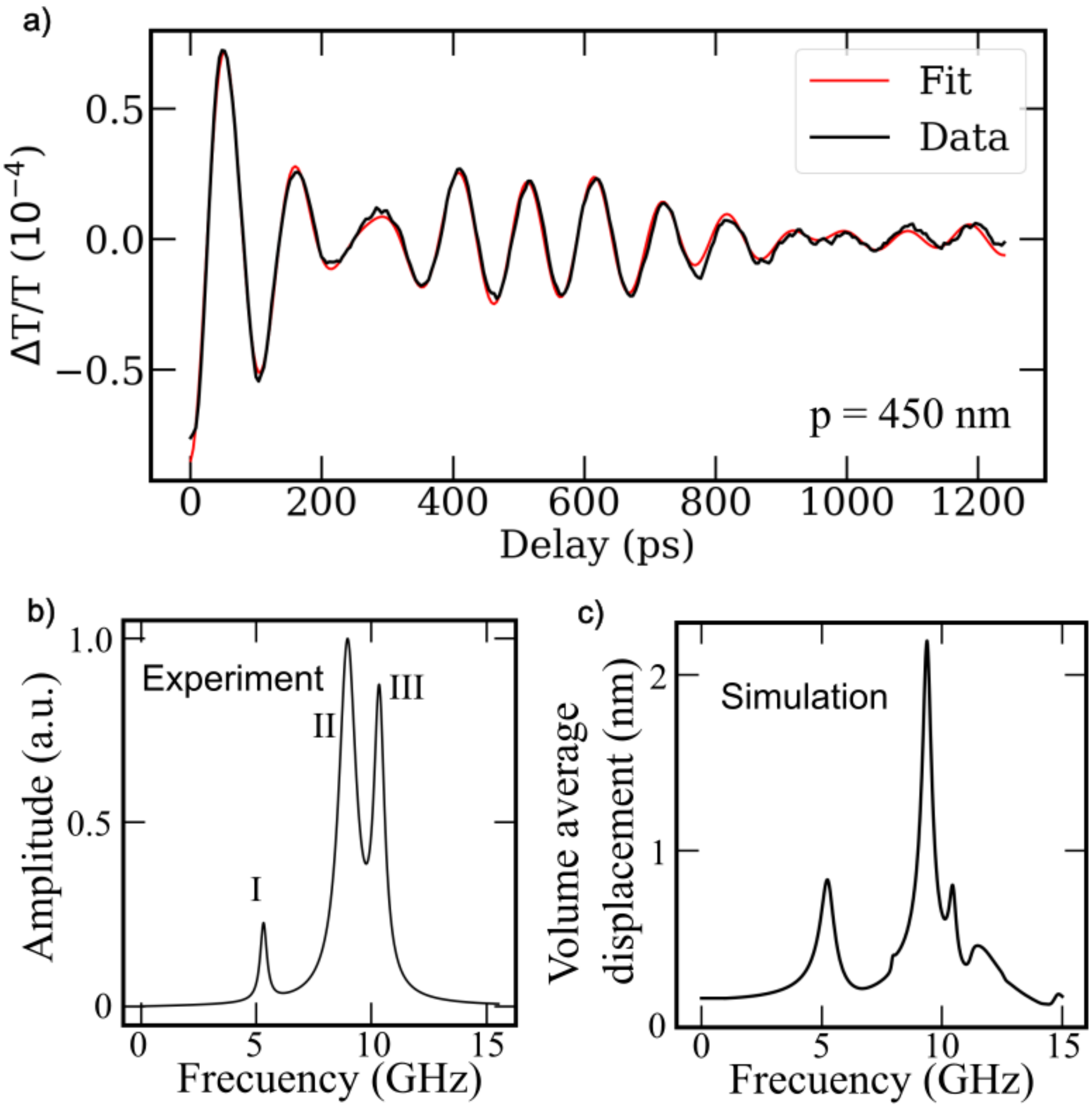

Figure 5. (a) Representative phonon signal and corresponding fit for a 450 nm-pitch plasmonic metasurface. (b) Mechanical spectrum obtained from the fit. The low-frequency mode I, already present in the isolated-disk simulation, becomes clearly visible in the metasurface response. (c) Simulated nanodisk displacement for the 450 nm-pitch metasurface.

As already observed for the isolated nanodisk, the relative amplitudes of the experimentally detected modes do not directly reproduce the calculated mechanical displacements, evidencing a mode-dependent optical transduction efficiency. This is consistent with the dependence of phonon–plasmon coupling on the spatial overlap between the optical field and the mechanical deformation, as discussed in Ref.[15]. The same selectivity remains evident in the metasurfaces. In particular, while the calculated relative amplitudes of modes II and III vary appreciably with lattice pitch, their experimental amplitude ratio remains comparatively stable over the 400–500 nm range (Fig. S1). The measured differential transmission should therefore not be interpreted as a direct measure of mechanical displacement, but rather as a mechanically driven signal weighted by a mode-dependent optical sensitivity.

Mechanical simulations for the 450 nm-pitch metasurface (Fig. 5c) nevertheless show that mode I acquires a significantly larger mechanical contribution than in the isolated nanodisk, consistent with the experimental spectrum. Together with the mode-dependent substrate coupling already identified for the isolated nanodisk in Fig. 4, this points to substrate-mediated interactions between neighboring nanodisks as an important contribution to the mechanical response of the metasurface. Each nanodisk, driven by the pump pulse, launches elastic waves into the substrate through its normal modes. These waves provide a mechanical interaction between neighboring nanodisks whose magnitude and phase depend on the lattice spacing and on the ability of each mode to radiate into the substrate.

Simulations of the nanodisk displacement for metasurfaces are presented in Fig. 6 as a function of frequency and lattice pitch, with the experimentally measured frequencies overlaid as white circles. The map shows how the three intrinsic nanodisk resonances identified in Fig. 3 evolve when the disks are periodically arranged. Their response to the periodic environment is strongly mode dependent, consistent with the surface acoustic energy fraction of Fig. 4c, which shows that modes I and III produce a substantially stronger elastic response in the substrate than mode II. This difference provides the key to understanding the different behaviors observed in Fig. 6.

A useful physical picture is provided by Kähler et al.,[39] who considered a simplified one-dimensional model of two resonators coupled through surface acoustic waves. Each resonator launches an elastic wave into the common substrate, which reaches the neighboring resonator with an amplitude that depends on their separation and a phase determined by propagation distance and frequency. The resulting substrate-mediated force can shift the resonance frequency and modify the radiative damping, with the two effects determined by the phase of the acoustic interaction relative to the resonator motion. Although this model is much simpler than the three-dimensional displacement fields of the present nanodisks, it captures the essential role of amplitude and phase in substrate-mediated mechanical coupling. In a periodic array, the same picture can be extended qualitatively to the combined contribution of several neighboring resonators, so that changing the pitch modifies the elastic interaction experienced by each nanodisk. Substrate-mediated frequency shifts with interparticle separation have also been observed directly in plasmonic nanodisk assemblies.[40]

Modes I and II represent two different limits of this interaction. Mode II remains predominantly confined to the nanodisk and generates comparatively little substrate motion. Its feedback through the substrate is therefore weak, and its resonance remains close to 9.4 GHz throughout the pitch range. Mode I, in contrast, couples much more efficiently to the substrate. The elastic field emitted by neighboring disks can therefore appreciably shift its resonance frequency. As the pitch is reduced, the changing propagation phase and amplitude of this feedback shift mode I to higher frequency. Its behavior is thus naturally interpreted as an off-resonant, dispersive manifestation of substrate-mediated coupling.

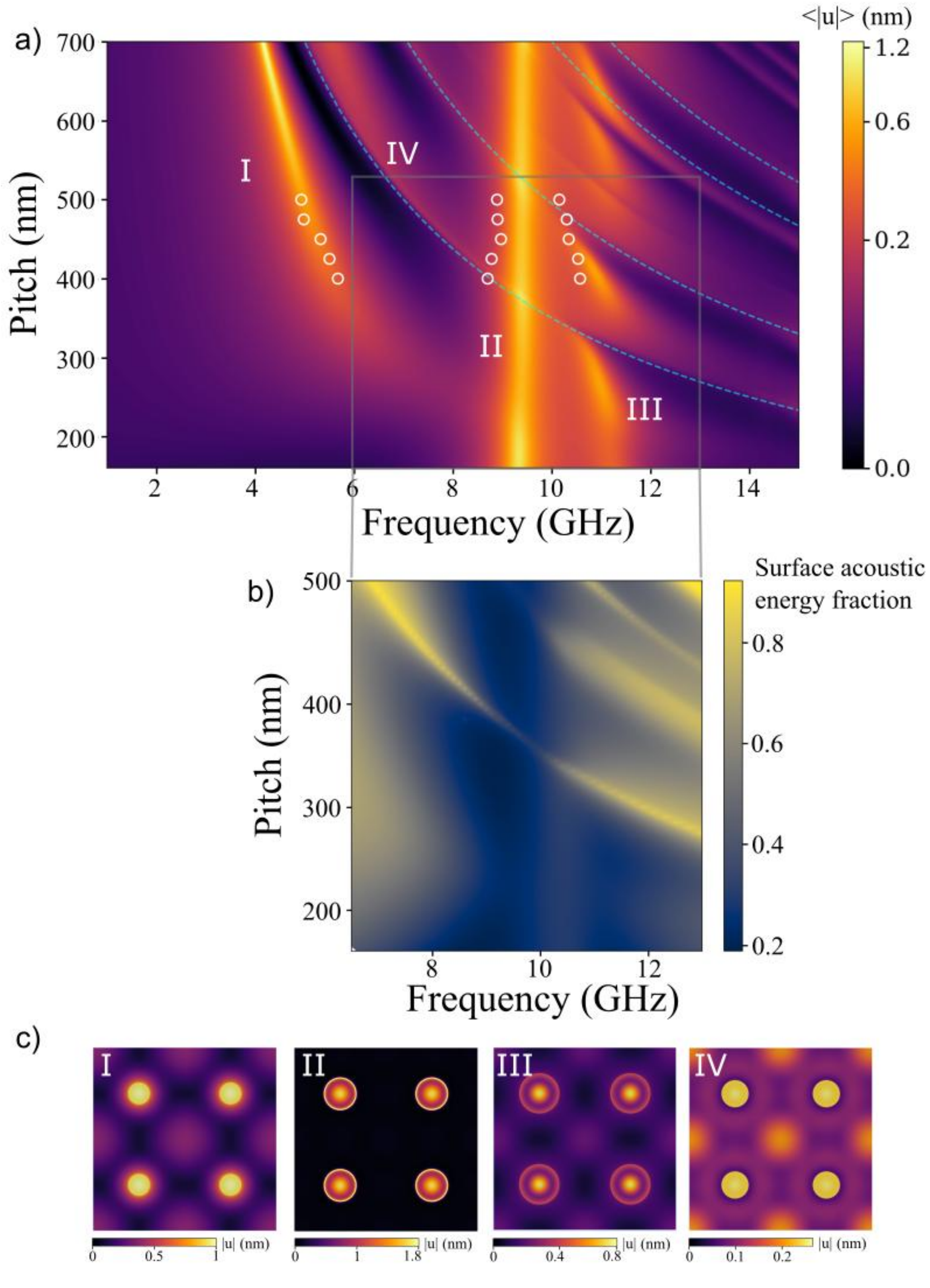

Figure 6. (a) Simulated nanodisk average displacement as a function of frequency and lattice pitch. Experimentally measured frequencies are overlaid as white circles. Turquoise dashed lines indicate the empty-acoustic-lattice reference branches. The map shows the nearly pitch-independent mode II, the strongly dispersive mode I, and the avoided crossing between mode III and the lattice-selected mode IV. (b) Surface acoustic energy fraction in the highlighted frequency-pitch region, showing the strong mode dependence of substrate coupling and the enhanced substrate contribution around the III-IV avoided crossing. (c) Representative displacement maps for modes I-IV, illustrating the different balance between nanodisk and substrate motion.

The periodic structure also supports lattice-selected acoustic branches associated with the reciprocal lattice of the array. For a square lattice and an approximately linear substrate acoustic dispersion, their uncoupled reference frequencies are $f_{ij} = \frac{v_{eff}}{p}\sqrt{i^2 + j^2}$ where $p$ is the lattice pitch, $i$ and $j$ are reciprocal-lattice indices, and $v_{eff} = 3500\ m/s$ is the effective phase velocity of the relevant substrate acoustic branch. Inverse-pitch dispersions with the corresponding sequence of reciprocal-lattice orders have been observed in time-resolved hypersonic phononic crystals.[18,41] These empty-acoustic-lattice curves provide a useful uncoupled reference. The actual metasurface modes may deviate from them because of mass loading and interaction with the nanodisks. Figure 6a shows for reference $f_{01}$, $f_{11}$, $f_{20}$, and $f_{21}$ branches, superimposed on the complete simulation (turquoise dashed lines).

Mode IV belongs to this lattice-selected class. Its interaction with mode III represents the resonant counterpart of the same substrate-mediated physics described above. Mode III, unlike mode II, produces appreciable substrate motion and therefore couples efficiently to the dispersive acoustic branch. Far from resonance the two modes retain predominantly nanodisk-like and substrate-like character. As their frequencies approach, the interaction becomes resonant, the eigenfrequencies repel, and the modes hybridize, producing the avoided crossing observed in Fig. 6. This is the characteristic local-resonance hybridization mechanism discussed for surface-resonator phononic structures.[42,43] Figure 6b shows the near-surface substrate participation in the region of the III–IV avoided crossing. A pronounced increase is observed around the crossing, indicating that the coupled modes involve a larger contribution from substrate motion where modes III and IV interact most strongly. This behavior is consistent with the change in modal character expected upon hybridization and supports the interpretation of the avoided crossing as arising from the coupling between the nanodisk resonance and the lattice-selected acoustic mode.

Figure 6c shows representative displacement maps for modes I–IV, illustrating the different relative contributions of nanodisk and substrate motion for each mode. The four modes can therefore be understood within a single physical picture. Mode II couples only weakly to the substrate and remains essentially intrinsic. Mode I couples efficiently to substrate motion but stays away from a resonant lattice crossing, leading mainly to a pitch-dependent frequency renormalization. Mode III also couples efficiently to the substrate and, when it approaches the lattice-selected mode IV, enters the resonant regime and hybridizes with it. The different evolution of the three intrinsic modes is therefore determined by the combination of lattice geometry and the mode-dependent ability of each nanodisk vibration to launch elastic motion into the substrate.

The experimental frequencies follow the same overall picture. Mode II remains nearly independent of pitch, whereas mode I displays the strongest dispersion, shifting from approximately 5.6 to 4.9 GHz as the pitch increases from 400 to 500 nm. Mode III remains resolved in the 10-11 GHz range and lies in the frequency region where the simulations reveal its interaction with mode IV. Overall, the calculations reproduce the main spectral features and their evolution with pitch.

# Conclusions

We have investigated how lattice periodicity reorganizes the optical and mechanical response of gold nanodisk metasurfaces on silica. Optically, varying the lattice pitch shifts the Rayleigh anomaly through the broad localized plasmon resonance and produces a strongly asymmetric collective response. For a 500 nm pitch, the probe wavelength lies on a rapidly varying spectral region, resulting in a pronounced wavelength dependence of the detected phonon signal.

The mechanical response is strongly mode dependent. The three intrinsic nanodisk resonances couple differently to elastic motion in the substrate. Mode II remains nearly insensitive to pitch, mode I undergoes a pronounced pitch-dependent frequency renormalization through substrate-mediated interactions, and mode III interacts resonantly with the lattice-selected acoustic mode IV, producing an avoided crossing and hybridization. The simulations reproduce the experimental evolution of the resonances and relate these different behaviors to the mode-dependent participation of the substrate.

Taken together, these results show that lattice pitch provides a common geometrical parameter to tailor both optical sensitivity and mechanical response. Because electromagnetic and elastic interactions depend differently on geometry, the same metasurface can combine a sensitive optical readout with mechanical excitations that remain mainly confined to the nanodisk, couple efficiently to the substrate, or become hybridized with lattice-selected acoustic modes. This provides a route to designing plasmonic metasurfaces in which the regions of strongest electromagnetic and mechanical response can be chosen either to overlap or to remain spatially separated according to the desired function.

# Experimental techniques

## Sample fabrication

Gold nanodisks were fabricated on a silica substrate by electron-beam lithography (EBL). A PMMA 950k A4 resist layer was spin-coated onto the sample and baked for 3 min at 180 °C, followed by a layer of Espacer 300Z . The resist was selectively exposed to the EBL (Raith Eline Plus) as designed and developed in a methyl isobutyl ketone and isopropyl alcohol (1:3) solution for 2 min. A 2 nm Cr adhesion layer followed by 35 nm of Au was then deposited by sputtering-beam evaporation (Bestec). Finally, lift-off was performed in Microposit Remover 1165 to remove the remaining resist and excess metal. Micrographs of the Au nanostructures were acquired using a scanning electron microscope (Carl Zeiss GeminiSem 560).

## Pump-probe measurements

The pump and probe beams were generated from a mode-locked Ti:sapphire laser operating at 802 nm central wavelength with a 95 MHz repetition rate and pulses of approximately 300 fs at the sample. The ~400 nm pump was obtained by second-harmonic generation in a BBO crystal and modulated with an acousto-optic modulator for lock-in detection. The ~800 nm probe was delayed using a calibrated delay line. Pump and probe average powers were 90 μW and 30 μW, respectively, with spot diameters of 2.2 μm and 1.6 μm. Because both spot diameters exceed the lattice pitch, several neighboring nanodisks are simultaneously excited and probed. The objective numerical aperture was 0.6.

Signals were systematically analyzed by removing the exponential background through a nonlinear fit and using a linear-prediction algorithm to extract the frequency spectrum together with the associated amplitudes and lifetimes.[44]

## Numerical Simulations

For the mechanical simulations, a frequency-domain study was implemented using the Structural Mechanics module of COMSOL Multiphysics 6.3. The isolated nanodisk was modeled using a 2D axisymmetric geometry. Metasurface calculations were performed with a 3D single unit cell containing one nanodisk and periodic boundary conditions in the in-plane directions. In both cases, a gold nanodisk on a $SiO_2$ substrate was simulated, with a perfectly matched layer at the bottom of the substrate to mimic an effectively semi-infinite medium.

The material parameters used for gold were Young's modulus E = 70 GPa, Poisson ratio ν = 0.43, and density ρ = 19.3 g/cm³. For $SiO_2$, the corresponding values were E = 70 GPa, ν = 0.17, and ρ = 2.35 g/cm³. Homogeneous excitation of the nanodisks was assumed, modeling pump-induced heating ($\Delta T = 100$ K) uniformly over the gold with a thermal expansion coefficient of 14 × $10^{-6}$ $K^{-1}$ and negligible thermal expansion of the substrate. A 5% homogeneous damping was applied to the gold nanodisks to account phenomenologically for internal losses. Frequencies from 1 to 20 GHz and pitches from 160 to 700 nm were investigated. For the near-surface substrate participation shown in Figs. 4c and 6b, the substrate integral was restricted to a depth of one acoustic wavelength.

For the electromagnetic simulations, the Wave Optics module was used. The geometry was analogous to the mechanical model, except that a 3D model was used for the isolated nanodisk to account for the linearly polarized optical excitation. The surrounding air region and an upper PML were also included. Over the 400–1200 nm spectral range, the refractive indices of air and $SiO_2$ were taken as 1 and 1.45 respectively, while the wavelength-dependent complex refractive index of gold was described using the Rakic model implemented in COMSOL.

For the metasurface optical simulations, a full-field formulation using ports was employed to simulate normal-incidence plane-wave excitation and to calculate transmittance, reflectance, and absorbance. For the isolated nanodisk, a scattered-field formulation was used with the background field given by an incident monochromatic plane wave. This allowed the background and scattered fields to be separated and the scattering and absorption cross-sections to be obtained by integrating the normal component of the Poynting vector over the scatterer surface and the electromagnetic absorption over the disk volume, respectively.

# Funding and acknowledgements

This work was partially supported by ANPCyT (PICT 2021 IA-363 and PICT 2021 GRF-TI-00349), UBACyT (Proyecto 20020220200078BA), and Red Federal de Alto Impacto (FFFLASH; CONVE-2023−10189190). A.V.B. acknowledges funding support from the Alexander von Humboldt Foundation through the 2023 Georg Forster Award. D.X. and E.C. acknowledge funding and support from the LMU-CSC program, the Deutsche Forschungsgemeinschaft (DFG) under Germany´s Excellence Strategy-EXC 2089/2-390776260, the Bavarian Program Solar Technologies Go Hybrid (SolTech) and the Center for NanoScience (CeNS).